\documentclass[aps,prb,twocolumn,showpacs,superscriptaddress]{revtex4-1}

\usepackage{graphicx}
\usepackage{color}
\usepackage{txfonts}

\begin{document}


\title{Magnetic order in pyrochlore iridate Nd$_2$Ir$_2$O$_7$ probed by muon spin relaxation}



\author{Hanjie Guo}
\email[]{hjguo@zju.edu.cn}
\affiliation{Department of Physics and State Key Laboratory of Silicon Materials, Zhejiang University, Hangzhou 310027, China}
\affiliation{Advanced Meson Science Laboratory, RIKEN, 2-1 Hirosawa, Wako, Saitama 351-0198, Japan}

\author{Kazuyuki Matsuhira}
\affiliation{Facuty of Engineering, Kyushu Institute of Technology, KitaKyushu 804-8550, Japan}

\author{Ikuto Kawasaki}
\affiliation{Advanced Meson Science Laboratory, RIKEN, 2-1 Hirosawa, Wako, Saitama 351-0198, Japan}

\author{Makoto Wakeshima}
\affiliation{Division of Chemistry, Graduate School of Science, Hokkaido University, Sapporo 060-0810, Japan}

\author{Yukio Hinatsu}
\affiliation{Division of Chemistry, Graduate School of Science, Hokkaido University, Sapporo 060-0810, Japan}

\author{Isao Watanabe}
\affiliation{Advanced Meson Science Laboratory, RIKEN, 2-1 Hirosawa, Wako, Saitama 351-0198, Japan}

\author{Zhu-an Xu}
\affiliation{Department of Physics and State Key Laboratory of Silicon Materials, Zhejiang University, Hangzhou 310027, China}



\date{\today}

\begin{abstract}
Muon-spin relaxation results on the pyrochlore iridate Nd$_2$Ir$_2$O$_7$ are reported. Spontaneous coherent muon-spin precession below the metal-insulator transition (MIT) temperature of about 33 K is observed, indicating the appearance of a long-ranged magnetic ordering of Ir$^{4+}$ moments. With further decrease in temperature, the internal field at the muon site increases again below about 9 K. The second increase of internal field suggests the ordering of Nd$^{3+}$ moments, which is consistent with a previous neutron experiment. Our results suggest that the MIT and magnetic ordering of Ir$^{4+}$ moments have a close relationship and that the large spin-orbit coupling of Ir 5\textit{d} electrons plays a key role for both MIT and the mechanism of the magnetic ordering in pyrochlore iridates in the insulting ground state.
\end{abstract}


\maketitle


Pyrochlore iridates provide a fertile playground to investigate novel topological phases based on the network of corner-sharing tetrahedra structure and the relatively large spin-orbit coupling (SOC) inherent in Ir 5\textit{d} electrons.\cite{Pesin-theory, Kargarian-theory, Wan_theory, William-theory, Yang-theory} The interplay between SOC and electron-electron correlations (\textit{U}) produces characteristic electronic states. A series of \textit{R}$_2$Ir$_2$O$_7$ (\textit{R}-227, \textit{R} = Nd-Ho) compounds exhibit metallic or semi-metallic behavior and undergo metal-insulator transitions (MIT's) at temperature $T_\mathrm{MI}$,\cite{Matsuhira_transport} while Pr-227 shows metallic behavior down to 0.3 K.\cite{Nakatsuji-Pr-227}

The MIT has been observed to be accompanied by a magnetic anomaly. The magnetic susceptibility exhibits a bifurcation in the field-cooled (FC) and zero-field-cooled (ZFC) conditions below $T_\mathrm{MI}$.\cite{Matsuhira_transport} A muon-spin relaxation ($\mu$SR) experiment on Eu-227\cite{Zhao_Eu} observed the muon-spin precession in the zero-field (ZF) condition below $T_\mathrm{MI}$, indicating that MIT is accompanied by a long-ranged magnetically ordered state. In the case of Y-227, Shapiro \textit{et al.}\cite{Neutron-Y} reported the similar bifurcation of the magnetic susceptibility at $\sim$ 155 K, although MIT is indistinct from the resistivity measurement. Disseler \textit{et al.} \cite{Disseler_Y} reported the bifurcation of the magnetic susceptibility in the same system at $\sim$ 190 K and the appearance of the spontaneous muon-spin precession from ZF-$\mu$SR below $\sim$ 150 K. A local spin-density approximation calculation including \textit{U} and SOC showed that the non-collinear all-in/all-out (pointing toward or outward the center of the tetrahedra) spin configuration is indispensable to reproduce the insulating phase, indicating a possible relationship between MIT and magnetically ordered state.\cite{Wan_theory}

In this study, we focus on Nd-227 which shows metallic behavior at high temperatures and undergoes a MIT at $T\mathrm{_{MI}}$ of about 33 K, which is the lowest in the series of \textit{R}-227, and the magnetic susceptibility shows the bifurcation below $T\mathrm{_{MI}}$ in ZFC and FC conditions.\cite{Matsuhira_transport} A neutron study on Nd-227 revealed an ordering of Nd$^{3+}$ moments below $T\mathrm{_{Nd}}$ = 15 $\pm$ 5 K, and pointed out the appearance of a hidden ordering of Ir$^{4+}$ moments below $T_\mathrm{MI}$ since the Kramers ground doublet of Nd$^{3+}$ begins to split below $T_\mathrm{MI}$.\cite{neutron} Disseler \textit{et al.} performed the ZF-$\mu$SR experiment on Nd-227 and observed the muon-spin precession below 8 K.\cite{previous-muSR} They claimed that this muon-spin precession is attributed to an ordering of Ir$^{4+}$ moments while Nd$^{3+}$ moments are still fluctuating. Accordingly, those controversial results encourage us to re-study the magnetic properties of Nd-227 especially below $T_\mathrm{MI}$.

In this paper, we present ZF and longitudinal-field (LF) $\mu$SR results on Nd-227. We use a high quality sample with the same grade of that used for the neutron scattering experiment.\cite{neutron} We observe the spontaneous muon-spin precession below $\sim$ 30 K which is almost the same as $T_\mathrm{MI}$ = 33 K. In addition, another increase of the internal field at the muon site is observed below about 9 K. This result suggests that Ir$^{4+}$ moments form a long-ranged ordered state below $T_\mathrm{MI}$ and Nd$^{3+}$ moments show another ordering below about 9 K. The smaller value of the internal field from the ordered Ir$^{4+}$ moments at the muon site suggests the reduction of Ir$^{4+}$ moments compared with the other pyrochlore iridates which have the insulating ground state. Our results suggest a close relationship between the MIT and magnetic transition in Nd-227 as suggested in other \textit{R}-227 systems such as Eu-227.

A polycrystalline sample of Nd-227 was synthesized by a standard solid-state reaction method.\cite{Matsuhira-2007, Matsuhira_transport} The sample quality can be checked by both  the X-ray diffraction (XRD) \cite{Matsuhira-2007} and MIT. Single phase was confirmed by XRD with narrow and sharp peaks. The sharp MIT at 33 K, which is the same as determined on the samples used for previous macroscopic measurement\cite{Matsuhira_transport} and neutron scattering experiment,\cite{neutron} was confirmed as well. Recently, Ueda {\it et al.} synthesized a high-quality sample by using a high-pressure solid-state reaction method and their obtained value of $T_\mathrm{MI}$ is almost the same as that of ours.\cite{Ueda_sample} $\mu$SR experiments were performed at the RIKEN-RAL Muon Facility at the Rutherford-Appleton Laboratory in the UK.\cite{Matsuzaki-facility} Spin-polarized pulsed muons were injected into the sample and decayed positrons ejected preferentially along the muon-spin direction were accumulated. The initial muon-spin polarization is in parallel with the beam line. Forward and backward counters are located in the upstream and downstream of the beam line. The asymmetry parameter of the muon-spin polarization is defined as $A(t)=[F(t)-\alpha B(t)]/[F(t)+ \alpha B(t)]$, where $F(t$) and $B(t)$ are the number of muon events counted by the forward and backward counters at time $t$, respectively. The $\alpha$ reflects the relative counting efficiency of the forward and backward counters. The time dependence of $A(t)$ ($\mu$SR time spectrum) was measured at different temperatures. The LF was applied along the initial muon-spin polarization.

\begin{figure}
\centering
\includegraphics[width=8cm]{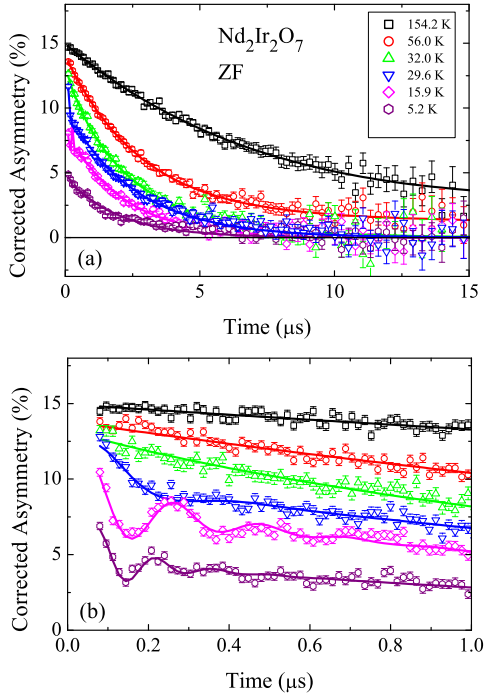}
\caption{(Color online) Zero-field $\mu$SR time spectra measured at various temperatures. Solid curves are fitting results by using Eq. (1). (a) shows the long-time muon-spin depolarization behavior up to 15 $\mu$sec; and (b) shows that in an early time region.}
\label{spectrum}
\end{figure}

Figure \ref{spectrum}(a) shows ZF-$\mu$SR time spectra at various temperatures. Spectra above $T_\mathrm{MI}$ show the exponential depolarization behavior. With decrease in temperature, the muon-spin depolarization becomes faster as the temperature is approaching to $T_\mathrm{MI}$. When the temperature is lower than $T_\mathrm{MI}$, a significant loss of the initial asymmetry at $t$=0, $A(0)$, is observed. Figure \ref{spectrum}(b) shows an early time region of the time spectrum. The time spectrum at 32 K does not show any remarkable change and no evidence of the appearance of a magnetically ordered state is obtained. Well-defined muon-spin precession develops with decreasing temperature below about 30 K. This observation gives a strong indication of the appearance of a long-ranged magnetically ordered state. The muon-spin precession becomes faster and its precession amplitude becomes smaller with decreasing temperature, and is not well-defined below about 5 K (data not shown).

\begin{figure}
\centering
\includegraphics[width=8cm]{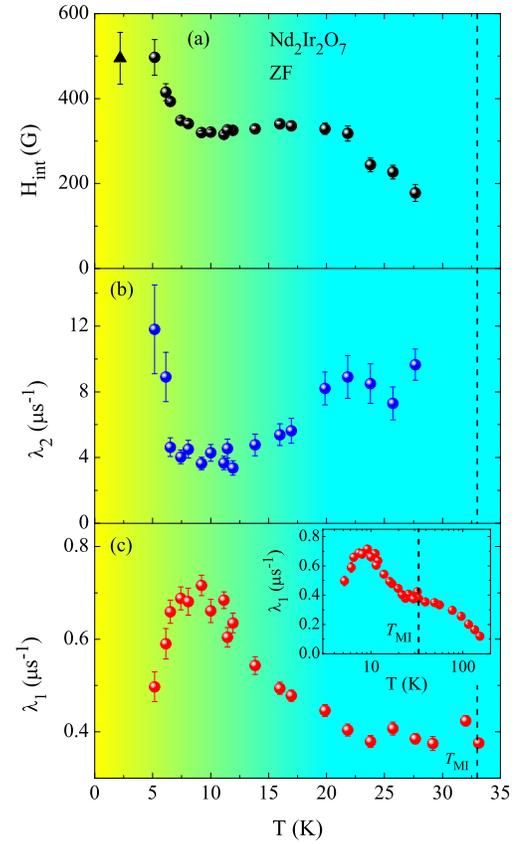}
\caption{(Color online) Temperature dependence of (a) the internal field at the muon site, $H_\mathrm{int}$; (b) the damping rate of the muon-spin precession, $\lambda_2$; and (c) the muon-spin depolarization rate, $\lambda_1$. The inset in (c) shows $\lambda_1$ in the whole measured temperature range. The broken line show the temperature of MIT. The yellow-colored area shows the temperature regions where the ordering of Nd$^{3+}$ moments occurs.\cite{neutron} The $H_\mathrm{int}$ at the base temperature in (a) was estimated by the LF measurement as described in the text.}
\label{field}
\end{figure}

In order to analyze the time spectra, the phenomenological function of
\begin{equation}
A(t)=A_1\mathrm{exp}(-\lambda_1 t)+A_2\mathrm{cos}(\gamma _\mu H_\mathrm{int}t+\phi)\mathrm{exp}(-\lambda _2 t)
\end{equation}
is used. The first and second terms express the exponentially depolarized and the muon-spin precession components with the initial asymmetries of $A_1$ and $A_2$, respectively. In the case of no muon-spin precession, only the first term is applied. Some muons are stopped at the silver sample holder making a background component. This background component has been subtracted from the analysis. Here, $\lambda_1$ represents the muon-spin depolarization rate of the first term, $\lambda_2$ and $\phi$ are the damping rate and the phase of the muon-spin precession, respectively. $\gamma _\mathrm{\mu}/2\pi$ = 13.55 MHz/kOe is the gyromagnetic ratio of the muon, and $H_\mathrm{int}$ is the internal field at the muon site. Principally speaking, Nd$^{3+}$ and Ir$^{4+}$ moments would give different contributions to $\lambda_1$. Since the spectra can be well fitted to Eq. (1), both contributions to the muon-spin depolarization can be described as the single-exponential function and it is hard to separate each contribution from one time spectrum. Fitted results are shown by solid curves in Fig. \ref{spectrum} and extracted parameters are plotted in Fig. \ref{field}.

Figure \ref{field}(a) shows the temperature dependence of $H_\mathrm{int}$. The $H_\mathrm{int}$ begins to increase with decreasing temperature below $T_\mathrm{MI}$. Since no clear muon-spin precession is observed at $T_\mathrm{MI}$ and a distinct change of the $\mu$SR time spectrum is observed below about 30 K, it is reasonable to identify that the magnetic ordering temperature is about 30 K from the present $\mu$SR measurement. This temperature is more or less similar to $T_\mathrm{MI}$ determined by the transport measurement. \cite{Matsuhira_transport} With further decreasing of temperature, $H_\mathrm{int}$ increases following the Brillouin type behavior and saturates below 20 K to the value of about 350 G, and increases again below about 9 K until the muon-spin precession becomes unobservable. Since $H_\mathrm{int}$ tends to exceed 500 G at lower temperatures, which is close to the limitation of the time resolution of the pulse muon facility,\cite{Matsuzaki-facility} the absence of muon-spin precession should be attributed to the limitation of the facility rather than the intrinsic disappearance of the long-rang ordered state.

Figure \ref{field}(b) shows the temperature dependence of $\lambda_2$, which reflects the damping rate of the muon-spin precession due to both dynamically fluctuating and statically distributed internal fields at the muon site.\cite{Adachi_lambda} Thus, the decrease in $\lambda_\mathrm 2$ in the magnetically ordered state indicates that magnetic moments become well aligned and/or that the spin fluctuations slow down. The $\lambda_2$ exhibits an upturn below about 9 K. Although absolute values of these data are unreliable due to the limitation of the time resolution of the pulsed muon beam, this change in $\lambda_2$ could be explained as that the distribution of $H_\mathrm{int}$ is enhanced and/or the spin dynamics becomes unstable again. Figure \ref{field}(c) shows the temperature dependence of $\lambda_1$ up to 35 K, while the data up to 150 K is depicted in the inset. This parameter can be regarded to reflect the dynamic fluctuations of the internal field at the muon site.\cite{Adachi_dynamic} Upon cooling from 150 K, $\lambda_1$ increases monotonically. An anomaly has been reported at about 120 K in the temperature dependence of $\lambda_1$ in the previous $\mu$SR experiment on Nd-227\cite{previous-muSR} and is attributed to the Ir$^{4+}$ magnetism. However, such an anomaly is not observable either in $\lambda_1$ or in the resistivity around 120 K in our measurements. With further decrease in temperature, $\lambda_1$ tends to level off around $T_\mathrm{MI}$ but increases again forming a peak at $\sim$ 9 K. No critical slowing down behavior is observed around 30 K where the long-ranged ordered state appears. The lack of the critical slowing down behavior has been also observed in Eu-227, Y-227 and Yb-227.\cite{Disseler_Y, Zhao_Eu}

\begin{figure}
\centering
\includegraphics[height=7cm]{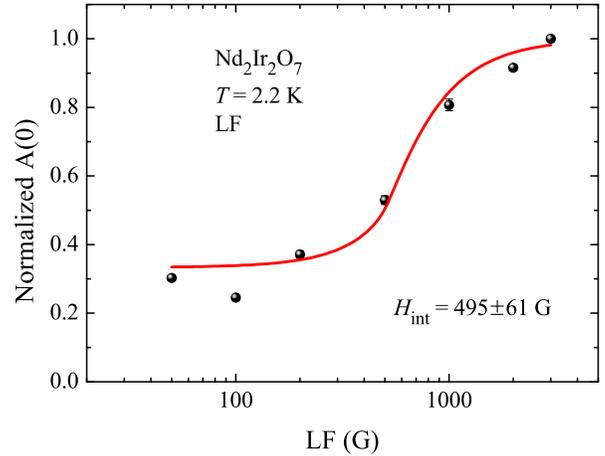}
\caption{(Color online) Longitudinal field (LF) dependence of the normalized initial asymmetry, $A(0)$ at 2.2 K. The red curve is the fit to Eq. (2). }
\label{LF}
\end{figure}

In order to obtain quantitative information of $H_\mathrm{int}$ at lower temperatures, we carried out LF measurements separately at 2.2 K. As performed in previous studies, $H_\mathrm{int}$ beyond the detectable limitation of the pulsed muon facility can be evaluated from the decoupling behavior of the muon-spin from $H_\mathrm{int}$ by LF.\cite{Watanabe_internal_field,Hachitani} Since fast and slow exponentially depolarized components are observed at 2.2 K in the LF (data not shown), time spectra in LF are analyzed by the phenomenological two-component function of $A(t)=A_{\mathrm{fast}}\mathrm{exp}(-\lambda_{\mathrm{fast}} t)+A_{\mathrm{slow}}\mathrm{exp}(-\lambda_{\mathrm{slow}} t)$, where $A_{\mathrm{fast}}$ and $A_{\mathrm{slow}}$ are the initial asymmetries of the fast and slow depolarized components, respectively. The $H_\mathrm{int}$ is evaluated by the LF dependence of the normalized initial asymmetry $A(0)$ by using the following function:\cite{Pratt_LF_formular}
\begin{equation}\label{asy_field}
    A(0)=\frac{3}{4}-\frac{1}{4x^2}+\frac{(x^2-1)^2}{16x^3}\mathrm{ln}\frac{(x+1)^2}{(x-1)^2},
\end{equation}
where $x = \mathrm{LF}/H_\mathrm{int}$. This equation assumes a unique internal field at the muon site with random orientation. The fitted result is the solid curve in Fig. \ref{LF}. The extracted $H_\mathrm{int}$ at 2.2 K is 495 $\pm$ 61 G and shown as a triangle mark in Fig. \ref{field}(a). The extracted value of $H_\mathrm{int}$ is closed to the direct evaluation at 5.2 K, suggesting that the internal field tends to saturate below about 5 K. We also note that this value is closed to that obtained by Disseler \textit{et al.}\cite{previous-muSR} at 1.6 K.

The neutron experiment suggested an ordering of Nd$^{3+}$ moments below 15 $\pm$ 5 K. \cite{neutron} Our $\mu$SR results show that the long-ranged magnetically ordered state appeares below $T_\mathrm{MI}$ and changes its ordered state again below $\sim$ 9 K. Taking into account the neutron experimental result, it is concluded that the change of the long-ranged ordered state observed in our study below 9 K is originated from the ordering of Nd$^{3+}$ moments. Accordingly, the increase of $\lambda_2$ and the peak of $\lambda_1$ around 9 K are regarded as the increase of the distribution of $H_\mathrm{int}$ and the slowing down behavior of Nd$^{3+}$ moments, respectively. The ordering temperature of Nd$^{3+}$ moments and the saturation of the internal field below $\sim$ 5 K are roughly consistent with the neutron result.\cite{neutron} As a result, the long-ranged ordered state below $T_\mathrm{MI}$ is concluded to be the one of Ir$^{4+}$ moments and its ordering temperature is similar to $T_\mathrm{MI}$. This result suggests that the long-ranged ordering of Ir$^{4+}$ moments in Nd-227 is closely related to the MIT as observed in Eu-227.\cite{Zhao_Eu} The magnetic volume fraction of the long-range ordered state can be estimated from the amplitude of the muon-spin precession and $A_1$, and at least 60\% of Ir$^{4+}$ moments show the long-range ordered state around 20 K. The magnetic volume fraction becomes almost full after Nd$^{3+}$ moments form the ordered state below 10 K.

\textcolor{red}{}

Following that the internal field at the muon site below $T_\mathrm{MI}$ should come from the long-ranged ordering of Ir$^{4+}$ moments, the saturated $H_\mathrm{int}$ of about 350 G below 20 K is much smaller than the other pyrochlore iridate with insulating ground state. In the case of Y-227, Yb-227 and Eu-227, the saturated $H_\mathrm{int}$ are 1100 G, 1100 G and 987 G, respectively. Suppose that the muon site in Nd-227 does not change drastically compared with these compounds and that the Ir sublattice has the similar magnetic structure, it is expected that the magnetic moment of Ir$^\mathrm{4+}$ in Nd-227 is reduced. The reduction of the Ir$^\mathrm{4+}$ moment may then suppress the magnetic ordering temperature in Nd-227. Indeed, the ordering temperatures in Eu-227, Y-227 and Yb-227 are 120 K, 150 K and 130 K, respectively, while it is only about 30 K for Nd-227. The reduction of the Ir$^{4+}$ moment could be caused by the larger hybridization between Ir 5{\it d} and O 2{\it p} orbitals in Nd-227 since Nd-227 is adjacent to Pr-227 which has the metallic ground state with no magnetically ordered state. Detailed numerical calculations of the ordered structure and magnitude of the Ir$^\mathrm{4+}$ moment are now going on including the estimation of hyperfine interactions at the expected muon site.

To reveal the mechanism of MIT and its relationship to the non-collinear all-in/all-out spin structure is a key issue to discuss novel electronic properties which are derived from strong SOC of Ir 5{\it d} electrons. Disseler \textit{et al.} suggested from their ZF-$\mu$SR study on Nd-227 that a disordered state of Ir$^{4+}$ moments appears below about 120 K where the resistivity shows an anomaly and that the disordered state changes into a long-range ordered one below about 8 K while the Nd$^{3+}$ moments keep paramagnetic down to 1.8 K.\cite{previous-muSR} Such a disordered intermediate state was also observed on Y-227 by the same group.\cite{Disseler_Y} They suggested that the Ruderman-Kittel-Kasuya-Yosida (RKKY) interaction between Ir$^{4+}$ moments induces the long-range magnetically ordered state. It should be noted that, the electronic property of their sample is different from that of the sample used in the neutron scattering experiment, showing unclear MIT and the less increase of the resistivity at low temperatures. In our study by using the high-quality sample, no anomaly from the resistivity and no sign of the disordered state are observed around 120 K. Furthermore, the system is insulating below $T_\mathrm{MI}$, which makes the RKKY interaction mechanism not to be suitable to explain the formation of magnetically ordered state and its relationship with the MIT.

Our results also reveal that the MIT and magnetic transition of Ir$^{4+}$ is strongly correlated. In the case of the Eu-227 system, a long-ranged magnetically ordered state of Ir$^{4+}$ moments appeares below $T_\mathrm{MI}$ although no critical slowing down behavior is observed around $T_\mathrm{MI}$. A change in the crystal structure is also observed in Eu-227 by a Raman scattering measurement to be accompanied with the long-ranged magnetically ordered state, whereas no detectable lattice distortion is observed for Nd-227.\cite{raman} This results in that the magnetic ordered state and the band gap at the Fermi surface are not induced by the releasing of geometrical frustration by means of lattice distortion.

Yamaura \textit{et al.} has suggested from the resonant X-ray scattering measurement on Cd$_2$Os$_2$O$_7$ that the Slater transition mechanism is excluded from possible origins of MIT because the all-in/all-out spin configuration does not break the lattice periodicity. Alternatively, they proposed a similar mechanism to the Lifshitz transition in which the non-collinear all-in/all-out spin structure causes the energy levels of the hole and electron bands moving downwards and upwards, respectively, inducing a gap at the Fermi level.\cite{Yamaura_Lifshitz_transition} Our present $\mu$SR study on Nd-227 reveals that the magnetic transition temperature of Ir$^{4+}$ moments is very close to $T_\mathrm{MI}$. It is quite difficult to say whether both temperatures are the same or not because each temperature is decided by different methods which have different characteristic time windows. Assuming that both temperatures are almost the same, the electronic state of Nd-227 can be regarded as the same with those of Eu-227 and Cd$_2$Os$_2$O$_7$. In this case, the Lifshitz-like transition which Yamaura {\it et al.} has suggested seems to be realistic for the understanding of the mechanism of MIT and the large SOC of Ir 5\textit{d} electrons should play a key role for MIT and the mechanism of the magnetic ordering. On the other hand, as observed in Y-227 and Yb-227 systems, the magnetically ordered state appears in the insulating state. If this is the case, the magnetic transition is not related to the MIT and the gap at the Fermi level is formed by different mechanisms like the Mott one as reported in Sr$_2$IrO$_4$.\cite{Kim-Sr2IrO4, Matsuhira_high_field} A clear model to explain about the mechanism of the appearance of the all-in/all-out spin structure and MIT is still unclear from this study. Careful and more detailed studies by using other experimental methods are needed around $T_\mathrm{MI}$ in order to clarify the relationship between the magnetic ordering and MIT.

In summary, ZF- and LF-$\mu$SR experiments were conducted to Nd-227 in order to investigate the magnetic properties below $T_\mathrm{MI}$ and its relationship with MIT. Long-ranged magnetically ordered states associated with Ir and Nd sublattices are observed below about 30 K and 9 K, respectively. The saturated internal field at the muon site transferred from the ordered Ir$^{4+}$ moments is reduced compared with other $R$-227 compounds with the insulating ground state, suggesting a reduction of the Ir$^{4+}$ moment due to the larger hybridization between Ir 5{\it d} and O 2{\it p} electronic orbitals in Nd-227. The magnetic transition of Ir$^{4+}$ moments is found to be closely related to the MIT. The band gap at the Fermi level can be formed by the similar mechanism of the Lifshiz transition triggered by the magnetic ordering with the non-collinear all-in/all-out spin structure showing the importance of the large SOC of Ir 5\textit{d} electrons, although there still be a possibility that the gap is induced by the Mott mechanism.

\begin{acknowledgments}
We thank Hui Xing and Yongkang Luo for useful discussions. This work was supported by a Grant-in-Aid for Scientific Research on Priority Areas "Novel Status of Matter Induced by Frustration (No. 19052005), Grant-in-Aid for Scientific Research on Innovation Areas "Heavy Electrons" (No. 21102518) and Grant-in-Aid for Scientific Research (C) (No. 23540417) and (B) (No. 23340096) from Ministry of Education, Culture, Sports, and Technology, Japan.
\end{acknowledgments}

\bibliography{reference}

\end{document}